\documentclass[11 pt,final,journal,letterpaper,oneside,onecolumn]{IEEEtran}
\usepackage{physics}
\usepackage{amsmath}
\usepackage{amsfonts}
\usepackage{mathtools}
\usepackage{amsfonts}
\usepackage{amssymb}
\usepackage{array}
\usepackage{makeidx}
\usepackage{csquotes}
\usepackage{graphicx}
\usepackage{cite}
\usepackage{xcolor}
\usepackage{subcaption}
\usepackage{graphicx}
\usepackage{multicol}
\usepackage{epstopdf}
\usepackage{multirow}
\usepackage{pifont}
\usepackage{subcaption}

\begin{document}
\title{\huge Intelligent and Secure Radio Environments for 6G Vehicular Aided HetNets: Key Opportunities and Challenges}
\author{Wali Ullah Khan, Muhammad Awais Javed, Sherali Zeadally, Eva Lagunas, Symeon Chatzinotas\thanks{This work was supported by Luxembourg National Research Fund (FNR) under the CORE project RISOTTI C20/IS/14773976.}}

\markboth{Submitted to IEEE}%
{Shell \MakeLowercase{\textit{et al.}}: Bare Demo of IEEEtran.cls for IEEE Journals} 

\maketitle
\begin{abstract}
Reconfigurable meta-surfaces are emerging as a novel and revolutionizing technology to enable intelligent wireless environments. Due to the low cost, improved efficiency, and passive nature of reflecting elements, it is becoming possible to program and control the wireless environment. Since wireless physical layer technologies can generally adapt to the wireless environment, their combination with reconfigurable surfaces and deep learning approaches can open new avenues for achieving secure 6G vehicular aided heterogeneous networks (HetNets). Motivated by these appealing advantages, this work provides an intelligent and secure radio environment (ISRE) paradigm for 6G vehicular aided HetNets. We present an overview of enabling technologies for ISRE-based 6G vehicular aided HetNets. We discuss features, design goals, and applications of such networks. Next, we outline new opportunities provided by ISRE-based 6G vehicular HetNets and we present a case study using the contextual bandit approach in terms of best IRS for secure communications. Finally, we discuss some future research opportunities. 
\end{abstract}
\begin{IEEEkeywords}
6G vehicular aided HetNet, intelligent communication, physical layer security, reconfigurable meta-surface.
\end{IEEEkeywords}
\IEEEpeerreviewmaketitle
\section{Introduction}
Sixth-generation (6G) vehicular aided heterogeneous networks (HetNets) have become an important topic in wireless communications and information technology \cite{9383089, 9378794}. It is well-known that 6G vehicular aided HetNets can enhance the network capacity and coverage by deploying a number of small-cells and road side unites operating in a larger macrocell network. These 6G vehicular aided HetNets are key drivers to meet the critical demand for security and reliability and support intelligent transportation system and large-scale Internet-of-things. 

Existing research contributions have widely investigated the energy and spectral efficiency aspects of HetNets while security has received little attention. Traditionally, the main difference between different tiers of vehicular aided HetNet pertains to the allocation of power and spectrum reuse. The small-cell base stations (BSs) generally serve the indoor users while the beamforming design and power allocation are used to suppress the inter-cell and mutual interference. The quality of service and network performance for small-cell users is generally degraded due to the power difference and low coverage of small-cell BS. Another contributing factor is the rapid signal attenuation due to strong fading and non-line of sight (NLOS) for indoor communications. Thus, there is a need for a paradigm shift to enable efficient communication for 6G vehicular aided HetNets.

At present, advanced transportation systems are designed to adapt according to the changes in the radio environments. The changes in the propagation channel are either leveraged or mitigated as per the application or service requirements. However, the physical objects that determine the propagation of electromagnetic waves through an environment are neither programmable nor controllable. In general, they are perceived by communications engineers as an obstacle to the whole communication process. This is because most of the transmitted energy by a transmitter to the receiver is either absorbed or scattered in the environment. Moreover, there are other limitations such as cost of deployment and selection of sites. In contrast, an intelligent radio environment is defined as the physical programmable space that processes information and plays an active role in the reliable exchange of data between a source and a destination. This is enabled through reconfigurable meta-surfaces and other deep learning (DL)-based data computation techniques for optimization, programmability, and controllability of the environment and are collectively known as intelligent reconfigurable surface (IRS) \cite{mirza_matching}. An IRS is a very thin sheet of electromagnetic material such that the arrangement of meta-atoms on this sheet can be controlled or programmed by deep learning in the presence of external stimuli \cite{irs_1}. Although the IRS-enabled antennas may ensure the reliable exchange of data, they would not be practical without any intelligent link security framework.

Protecting the security of information is becoming increasingly difficult due to the integration of all sorts of technologies in wireless environments. This is especially important for vehicular aided 6G HetNets because of the utilization of shared spectrum which make the information easy to intercept and expose it to malicious attacks. Existing networks mainly rely on higher layer encryption techniques whose effectiveness mainly depend on the computation capabilities of eavesdroppers. An eavesdropper with a relatively large computation power could solve the mathematical problem associated with the encryption technique. The management of cryptographic keys is not simple in vehicular aided HetNets with multi-tier communications. To complement the functionality of these cryptographic techniques, physical layer security (PLS) has been proposed as a viable solution that makes use of different characteristics of the wireless channel such as noise, fading, and interference. Different PLS techniques allow devices to exchange information without any signal leakage which is a desirable trait of 6G vehicular aided HetNets. In fact, the future success and rapid adoption of such networks heavily depends on the ability of devices to communicate securely and seamlessly without worsening the interference footprint or incurring any additional costs.

\begin{figure*}[!t]
\centering
\includegraphics[width=0.8\textwidth]{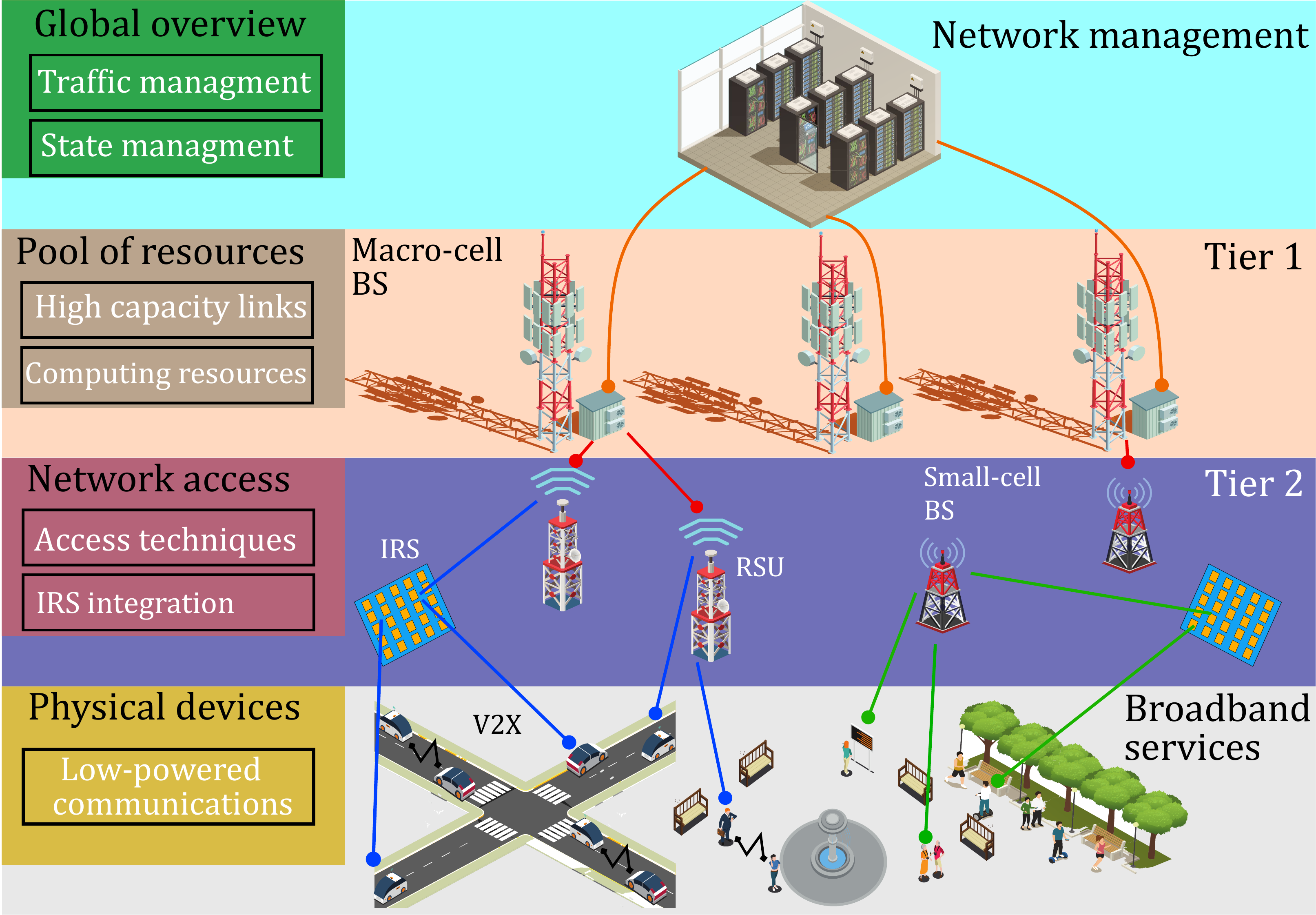}
\caption{An ISRE-based 6G vehicular aided HetNet. Network management is performed based on pre-defined policies for providing different services. Tier 1 consists of macro-cells which operate as a traditional cellular communication network and commonly use high-power for data transmissions. The tier 2 communication infrastructure mainly consists of small-cells and roadside unites. The transmission power in this region is lower and therefore has a lower coverage area. Different wireless services are provided to communication devices and vehicles that access the network through small-cells and roadside unites. Here, BS: Base Station, IRS: Intelligent Reflecting Surface, RSU: Road Side Unit, V2X: Vehicle to Everything.}
\label{block1}
\end{figure*}

The secrecy performance of 6G vehicular aided HetNets can be improved further by not only optimizing the design of the transmitter or receiver but also programming and managing the resources in the environment. To enable programmable radio environments, this work combines the emerging vision of intelligent and secure radio environments (ISRE) with 6G vehicular aided HetNets. Fig. \ref{block1} depicts the ISRE-based 6G vehicular aided HetNets. 

\textbf{Research contributions of this work:} 

We propose an efficient design approach for 6G vehicular aided HetNets that improves the reliability, security and energy efficiency. We summarize the main research contributions of this work as follows:

\begin{itemize}

\item The low-powered devices in 6G vehicular aided HetNets may not be equipped with multiple antennas and may require covert communications. For this reason, we need to ensure friendly jamming capabilities in 6G vehicular aided HetNets. We combine deep learning and reconfigurable meta-surfaces and provide a cost-effective solution as compared to conventional phased array antennas for jamming the reception of eavesdroppers.
\item The opportunity to program and control the resources in the wireless environment provides robust optimization techniques for improving the security of 6G vehicular aided HetNets. ISRE-based 6G vehicular aided HetNets reduces the energy scattering in unwanted directions by directing it toward the desired users.
\item ISRE-based 6G vehicular aided HetNets also reduce the computational complexity by employing passive IRS antennas. The low computation complexity not only reduces the energy footprint of communication devices and vehicles but also minimizes the computation delays.
\item Due to the passive processing of the reconfigurable meta-surfaces, ISRE-based 6G vehicular aided HetNets mitigate self-interference. This provides an added advantage over conventional full-duplex relays in improving the security of the 6G vehicular aided HetNets.

\end{itemize}

The remainder of the article discusses enablers of ISRE-based 6G vehicular aided HetNets. Then, we discuss the engineering aspects of ISRE-based 6G vehicular aided HetNets focusing on features, design goals, and applications. Next, we identify some future opportunities for new business models and improved user satisfaction followed by a case study and some research challenges. Finally, we make some concluding remarks.
  
\section{Enabling ISRE for 6G vehicular aided HetNets: Background and Overview}
In this section, we briefly describe the operations of three enablers (i.e., reconfigurable meta-surfaces, DL controllers, and PLS) for enabling secure and intelligent radio environments for 6G vehicular aided HetNets. 
\subsection{Reconfigurable Meta-surface}
One of the critical components of ISRE-based 6G vehicular aided HetNets is reconfigurable meta-surfaces. As we have mentioned earlier, the reconfigurable meta-surfaces can control the radio waves and are made up of electromagnetic materials \cite{irs_1}. The meta-surface is formed by a sub-wavelength dielectric or metallic scattering particles. One of the most important capabilities of such meta-surfaces lies in their ability to shape radio waves through active/passive beamforming \cite{9048622}. However, not all meta-surfaces are created the same and their reconfigurability largely depends on of the structure of their atoms. The meta-surfaces in which the meta-atoms have fixed structure cannot be reconfigured after their manufacture. For reconfigurable meta-surfaces, the arrangement of meta-atoms can be programmed in the presence of external stimuli. It is also worth pointing out that static meta-surfaces do not consume power whereas reconfigurable meta-surfaces operate in the semi-active mode for operating control switches \cite{mirza_matching}. 
The use of reconfigurable meta-surface provides more degrees of freedom over conventional communication techniques. It is well-known that, under practical settings, it is very difficult to optimize radio environments to improve the communication. Using the meta-surfaces, along with efficient deep learning approaches, the radio environments can be optimized jointly with the operations of devices at the source and the destination \cite{lit_2}. The reconfigurability of large meta-surfaces allows communication engineers to design dynamic models by using information about the channel conditions \cite{9048622}. As a result, the radio environment is not viewed as a random and uncontrollable entity, but as a critical component of the network itself that could be optimized to ensure the reliability and security of information.
\begin{table*}[!htp]
\caption{Comparison of different frameworks and their utility. In this table, \ding{76} refers to low, \ding{76}\ding{76} refers to average, and \ding{76}\ding{76}\ding{76} refers to high.}
\label{my-table}
\centering
\begin{tabular}{|p{2cm}|p{2cm}|p{1.5cm}|p{1.5cm}|p{1.5cm}|p{1.5cm}|p{1.5cm}|p{1.5cm}|}
\hline
\textbf{Framework} & \textbf{Language}       & \textbf{CNN $\ \ $capability} & \textbf{RNN $\ \ $capability} & \textbf{Speed} & \textbf{Training material and developer community} & \textbf{Architecture and ease-of-use} & \textbf{GPU support} \\ \hline
Theano             & C++, Python             & \ding{76}\ding{76}                      & \ding{76}\ding{76}                      & \ding{76}\ding{76}             & \ding{76}\ding{76}                                                 & \ding{76}                                     & \ding{76}                    \\ \hline
PyTorch            & Python, Lua             & \ding{76}\ding{76}\ding{76}                     & \ding{76}\ding{76}                      & \ding{76}\ding{76}\ding{76}            & \ding{76}                                                  & \ding{76}\ding{76}                                    & \ding{76}\ding{76}                   \\ \hline
Tensorflow         & Python                  & \ding{76}\ding{76}\ding{76}                     & \ding{76}\ding{76}                      & \ding{76}\ding{76}\ding{76}            & \ding{76}\ding{76}\ding{76}                                                & \ding{76}\ding{76}\ding{76}                                   & \ding{76}\ding{76}                   \\ \hline
MXNet              & Scala, Julia, R, Python & \ding{76}\ding{76}                      & \ding{76}                       & \ding{76}\ding{76}             & \ding{76}\ding{76}                                                 & \ding{76}\ding{76}                                    & \ding{76}\ding{76}\ding{76}                  \\ \hline
Caffe              & C++                     & \ding{76}\ding{76}                      & \ding{76}                       & \ding{76}              & \ding{76}                                                  & \ding{76}                                     & \ding{76}                    \\ \hline
CNTK               & C++                     & \ding{76}                       & \ding{76}\ding{76}\ding{76}                     & \ding{76}\ding{76}             & \ding{76}                                                  & \ding{76}                                     & \ding{76}                    \\ \hline
Neon               & Python                  & \ding{76}\ding{76}                      & \ding{76}                       & \ding{76}\ding{76}             & \ding{76}                                                  & \ding{76}                                     & \ding{76}                    \\ \hline
\end{tabular}
\end{table*}
\subsection{Deep learning-based Controller}
The ISRE-based 6G vehicular aided HetNets needs a deep learning-based controller located at the BS to optimize the radio environment and transform a reconfigurable meta-surface into an intelligent reflecting surface \cite{9526766}. There are many characteristics of the IRS that must be controlled by some deep learning approach. Some of these include the service-related data request by the users in the network, the relative positions of meta-surfaces in the environment, the location of eavesdropper and friendly jammers, and the mobility of different users in the network. Deep learning approaches are already gaining significant attention and they can play a vital role in securing 6G vehicular aided HetNets\cite{mao2018deep}. Significant advances have been made in other areas by well-known companies, e.g., Natural Language Processing (NLP) in Alexa, Computer vision in Facebook, and self-driving cars all of which have made considerable investments in these areas. However, in this work, we focus on the applications of deep learning in wireless networks and not on conventional pattern recognition approaches \cite{9812512}. Thus, we briefly summarize different types of neural networks along with their key characteristics. Based on learning, there are three major categories of neural networks.

\subsubsection{Supervised Learning}
This is the most generic form of deep learning, where the data is carefully labeled to support classification and clustering tasks. The simplest feed-forward artificial neural networks utilize the labeled data to perform the training. Another commonly used neural network employing supervised learning is a convolutional neural network (CNN). Their popularity has recently grown due to their ability to perform exceptionally well with images. Recurrent neural networks (RNN) and long short-term memory (LSTM) networks also fall in the category of supervised learning. LSTM is mainly used for the analysis of time-series data. Table \ref{my-table} presents some of the key frameworks for implementing deep learning models.

\subsubsection{Unsupervised Learning}
Unsupervised learning makes use of semi-/unlabeled data to make predictions and recommendations. This type of learning becomes important when neural networks handle a large amount of data and must analyze and predict certain aspects of the data. Key examples of these types of neural networks are self-organizing maps (used for feature detection), deep Boltzmann machines (used for recommendations), and auto-encoders and generative adversarial networks (used for generating data).

\subsubsection{Deep Reinforcement Learning}
\begin{figure*}[!t]
\centering
\includegraphics[width=0.8\textwidth]{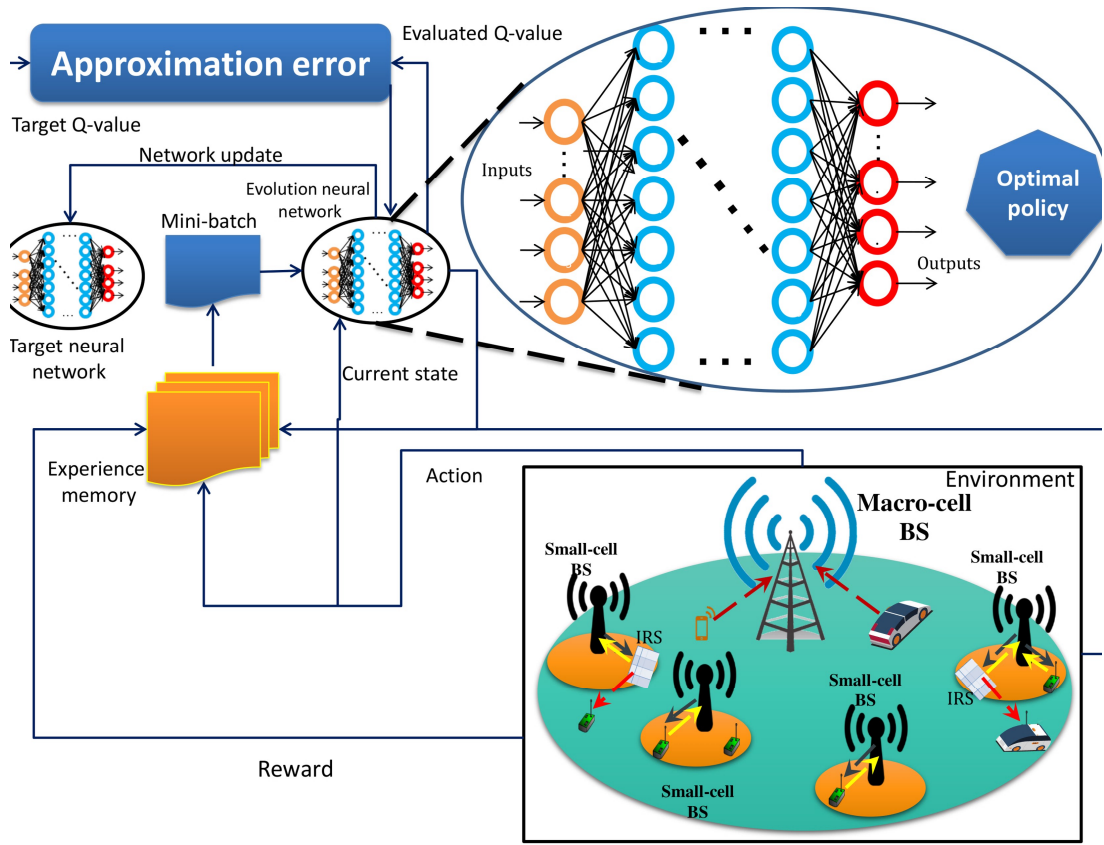}
\caption{An illustration of deep Q-learning model for the ISRE-based 6G vehicular aided HetNet environment. The model uses a deep neural network to find the approximate Q-values. The model includes experience replay mechanism (to remove the correlation between different observations), a feature set (given as input to the deep neural network), a target Q-network
for updating the primary Q-network and the simulation environment for extracting different parameters.}
\label{figQ}
\end{figure*}

Deep reinforcement learning refers to the type of learning where a policy/ value function is approximated using the reinforcement method. As the name suggests, the neural network (an agent) learns through reinforcement by interacting with the environment. The major goal of the network is to optimize the actions, taken at each step of learning, to achieve the best outcome. Perhaps, due to this flexibility in the interaction, there are many recent studies on IRS that make use of deep reinforcement learning. These studies consider many aspects from modulation and coding for the timely scheduling of data, and optimal spectrum access. There are many variants of deep reinforcement learning and some of the commonly used are deep contextual bandit (CB) \cite{8897638}, distributed proximal policy optimization, and deep policy gradient \cite{mao2018deep}. To better understand the operation of these techniques, Fig. \ref{figQ} shows the implementation of a fully connected deep Q-learning model.
\subsection{Physical Layer Security}
The intelligent radio environments improve the reliability of the wireless network. However, without proper security mechanisms to support reliability, such a network would be susceptible to attacks by the malicious users. Thus, PLS techniques are important to achieve ISRE-based 6G vehicular aided HetNets. This is also because broadcasting and superposition properties of the wireless medium make the network susceptible to security attacks \cite{pls_1}. There are PLS techniques that may help in improving the secrecy performance of wireless networks. Most of the PLS techniques either use relaying, jamming or a combination of both as we describe below:
\subsubsection{Cooperative Relaying}
PLS techniques use a trusted relay to forward the message from a source to a destination. Some of the common approaches include secure beamforming, efficient power allocation, relay selection, and relay ordering \cite{chu2019intelligent}. However, finding a trustworthy relay is a key challenge for cooperative relaying techniques.
\subsubsection{Friendly Jamming}
When intermediate relays are not available to improve the capacity of the desired wireless link, then friendly jammers can be used to confuse the eavesdropper. These jammers produce artificial noise to jam the reception of eavesdropper thereby hampering its ability to decode the signal. Artificial noise beamforming, incentive-based jamming, and partial jamming are some of the well-known approaches \cite{pls_1}. 
To establish a secure propagation environment for 6G vehicular aided HetNets, we need to combine the capabilities of PLS and IRS. In this context, a recent study proposed optimal beamforming solutions \cite{mirza_matching} while another study \cite{shen2019secrecy} leveraged non-convex modulus constraints to improve the secrecy performance. Another study \cite{9526766} provided security against multiple eavesdroppers using a deep learning technique. These studies, along with many others, have demonstrated the feasibility of combining PLS and IRS for developing a secure and intelligent propagation environment \cite{mirza_matching,shen2019secrecy,9526766}.
\section{Engineering ISRE-based 6G vehicular aided HetNets: Features, Design Goals, and Applications}
In this section, we discuss some of the required features, design goals, and applications of ISRE for 6G vehicular aided HetNets. 
\subsection{Features of ISRE-based 6G vehicular aided HetNets}
Next, we identify some key features of ISRE-based 6G vehicular aided HetNets that will pave the way for their practical implementations.
\subsubsection{Openness} 
Beyond gathering and presenting raw data for specialized services, the ISRE-based 6G vehicular aided HetNets must be flexible enough for different 3rd party applications. This is especially important for next-generation transportation systems where safety applications may require high security and reliability for the provisioning of timely services \cite{javed_cv2x_2020}.
\subsubsection{Multiwave functionality} 
The ISRE-based 6G vehicular aided HetNets can be exposed to different eavesdropping conditions. Therefore, the IRS must be able to achieve different functions which may include blocking certain radio waves, refracting the impinging signal or completely reflecting it some other times \cite{irs_1}.
\subsubsection{Resilience and reliability} 
The ISRE-based 6G vehicular aided HetNets must be resilient enough to guarantee a specific level of availability. Additionally, these networks must be able to provide highly reliable services based on the requirements of the different applications. 
\subsubsection{Configurability and interoperability} Configurability is a key feature of ISRE-based 6G vehicular aided HetNets, but it also comes at the cost of increased complexity. Nonetheless, to adjust according to the dynamic nature of 6G vehicular aided HetNets and to ensure semantic interoperability among different tiers of HetNet, different components of ISRE must be able to interact with each other seamlessly.
\subsubsection{Data Manageability} 
Different elements of ISRE-based 6G vehicular aided HetNets may produce a different type of data that which may include sensing, control, or video data. We need to clearly define the secure management policies regarding which type of data that could be stored and accessed by different entities of the network.
\subsection{Design Goals}
Next, we present some specific design goals for engineering ISRE-based 6G vehicular aided HetNets. The design goals for these networks comprise system metrics to achieve the desired level of system efficiency and secrecy performance. These design goals can be generally divided into two classes, i.e., services and resources.
\subsubsection{Service goals}
Service goals indicate the level of service satisfaction because 6G vehicular aided HetNets require guarantees for a range of service parameters in operational conditions. There are four metrics for ISRE, i.e., secrecy rate, network delay, outage probability, and network coverage \cite{shen2019secrecy}. Each of these service goals focuses on the needs of users in the network. Different users may have different goals and based on their needs, the priority of applying these goals may vary significantly.
\subsubsection{Resource goals}
The service goals alone are not enough to design an efficient wireless network. We need to specify wireless resources along with the cost of exchanging data and build reliable connections. These goals would be useful in regulating and identifying the measures for network usage. In this context, the manageable resources in HetNets are characterized by four different aspects, i.e., spectrum, time, power, and computing.
\subsection{Applications}
Although there can be many applications and use cases of the ISRE-based wireless network, this section presents some of the key applications of ISRE in the context of 6G vehicular aided HetNets.
\subsubsection{Passive Beamforming} 
The reconfigurable meta-surfaces, along with efficient deep learning approaches, can perform passive beamforming in a nearly batteryless manner \cite{shen2019secrecy}. This function significantly improves the signal power received at the receiver thereby improving the overall achievable secrecy rate. There are three types of tunable beamforming functions that the ISRE can perform. This includes beam steering, beam splitting, and guided radiation. Beam steering can be used to direct the impinging beam from different transmitters to increase the signal power of useful signals toward the intended receiver. The same function can be used to direct the jamming signal toward the eavesdroppers. Beam splitting is useful for deflecting the beam into different beams while guided radiation allows temporal-spatial distribution of the power of impinging radio waves. 
\subsubsection{Secure Multipurpose Surface} 
The ISRE can make use of multipurpose functionalities of a reconfigurable meta-surface. This meta-surface, along with artificial neural networks,  can be configured to reflect or refract the impinging signal from the source toward the destination and away from eavesdroppers \cite{9526766}. We note that, generally, PLS techniques are heavily dependent on fading conditions and the location of the receiver to prevent leakage of information. This aspect of ISRE makes it more suitable for dense 6G vehicular aided HetNet where the legitimate receiver could be located anywhere in the environment.   
\subsubsection{Efficient Encoding} 
Similar to typical backscatter communications, the ISRE itself can be used as a data generation platform to support ongoing secure communications in 6G vehicular aided HetNets. Using appropriate learning models, the meta-atoms can be modulated to perform secure communication to the receiver. Since the reconfigurable meta-surface does not generate any signal of its own and modulate the ambient radio signal \cite{irs_1}, it would be very difficult for eavesdroppers to decode such complex signals. However, we expect that such operations would consume more energy thereby requiring efficient power allocation strategies.
\subsubsection{Signal Manipulation} 
The combination of deep learning approaches and reconfigurable meta-surfaces can also be used in various ways to improve the secrecy performance of 6G vehicular aided HetNets. In this context, deep learning-based predictive solutions along with meta-surface can be quite useful. For instance, carefully trained RNN can be used to predict channel variations. Similarly, a deep reinforcement learning framework (as Fig. \ref{figQ} shows) can be efficiently used for symbol detection thereby avoiding the need for estimating the channel state. As a key principle, the ISRE uses existing radio signals for secure communications. However, an important component of this process is the appropriate selection of the IRS due to the path loss involved between a source and a meta-surface. Appropriate learning approaches are important for such techniques for the optimization of the entire 6G vehicular aided HetNet. In the next section, we consider a similar problem as a case study.
\section{Opportunities for ISRE-based 6G vehicular aided HetNets}
The emergence of ISRE and its integration with 6G HetNets opens up unprecedented opportunities. This section discusses some of these opportunities.
\subsection{New Business Models}
Construction companies and house owners in collaboration with telecommunication providers can make use of the meta-surfaces to build new business models and generate new revenue streams. It is likely that ISRE-based 6G vehicular aided HetNets would alter the way some businesses and consumers approach the world and surrounding environments. Accordingly, the business would require introducing new services to explore the ever-changing and ultra-connected landscape in the 6G era \cite{6g_h3}.  
\subsection{Cloud-based Cost-effective Applications}
Secure, flexible, and cost-effective cloud-based applications can be helpful in transforming the ISREs into an efficient decision-making platform. Such platforms can gather data from the environment and transfer it to a cloud for analysis. The cloud servers can analyze different parameters of the data and predict the changes in the environment for future references.
\subsection{Improved End-User Satisfaction}
Due to programmable smart environments and the real-time availability of information in ISRE-based 6G vehicular aided HetNets, we expect that the services provided to end-users will improve significantly. The smart environments can rapidly adapt based on the requirement of end-users and improve the network efficiency, resulting in higher end-user satisfaction.
\subsection{Intelligent Management Solutions}
The development of smart environments would result in the exponential growth of data to meet the ever-changing demands of users. This data would allow operators to design intelligent management solutions and policies depending on the traffic requirements of the network. Such management policies would considerably help ISRE-based 6G HetNets to outperform their conventional counterparts.
\subsection{Continuous Sensing} 
One of the key advantages of ISRE-based 6G vehicular aided HetNets is their ability to continuously sense the medium. Generally, a large amount of energy and time is consumed by channel estimation and sensing. This generally consumes a lot of power and involves signal processing computation, and information exchange in HetNets. With passive IRSs, the overall cost of channel sensing becomes negligible thereby improving the resource utilization of 6G vehicular aided HetNets.
\subsection{Faster Topological Convergence}
Smart environments are expected to enable faster topological convergence through the dissemination of signaling information. This affects the routing protocols for 6G vehicular aided HetNets that may rely on building a topology tree. This feature would also be helpful in improving the performance of the flooding-based protocols.
\section{ISRE-based 6G vehicular aided HetNets: A Case study}

\begin{figure*}[t]
     \centering
     \begin{subfigure}[b]{0.45\textwidth}
         \centering
         \includegraphics[width=\textwidth]{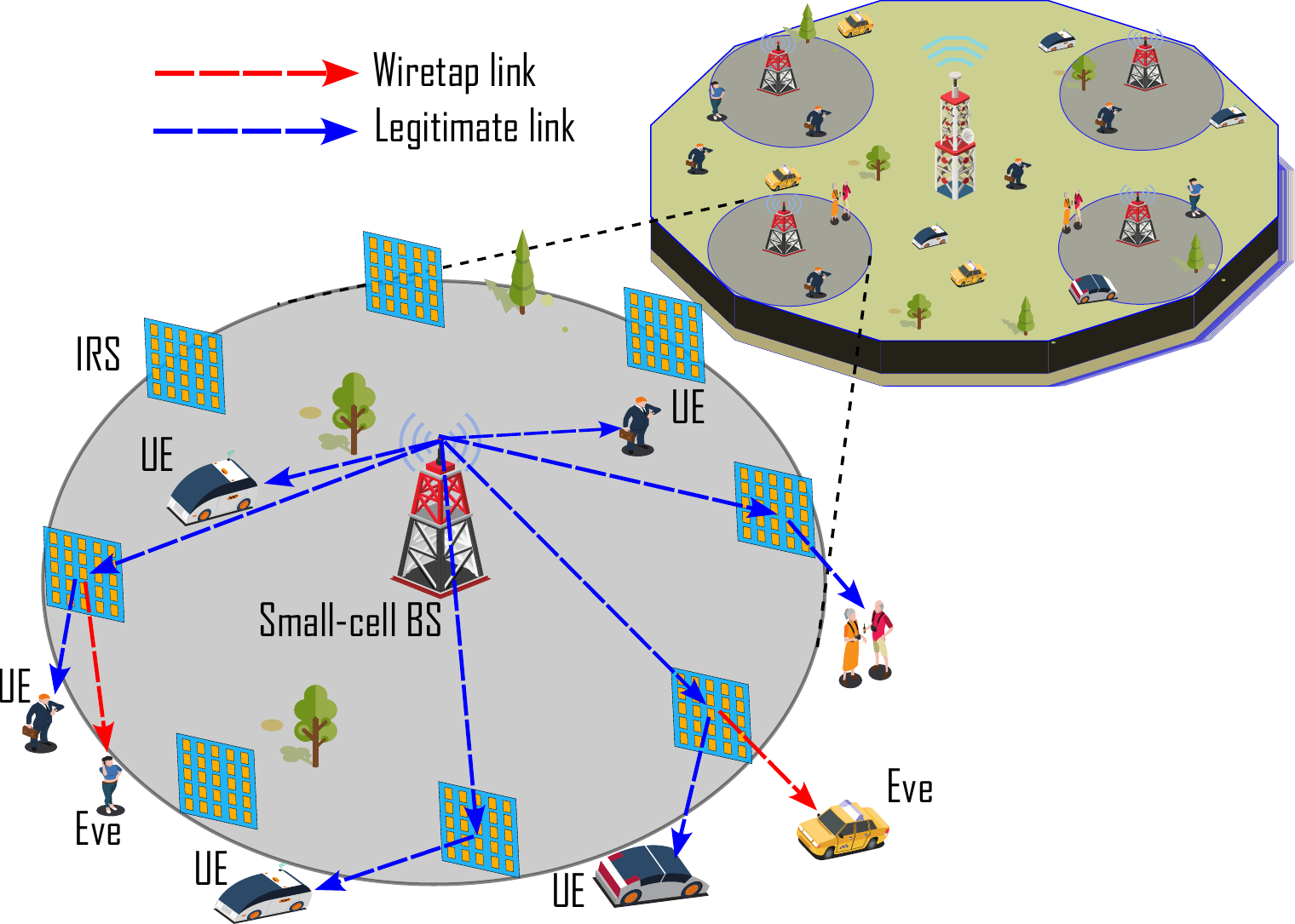}
         \caption{}
         \label{fig:1}
     \end{subfigure}
     \begin{subfigure}[b]{0.45\textwidth}
         \centering
         \includegraphics[width=\textwidth]{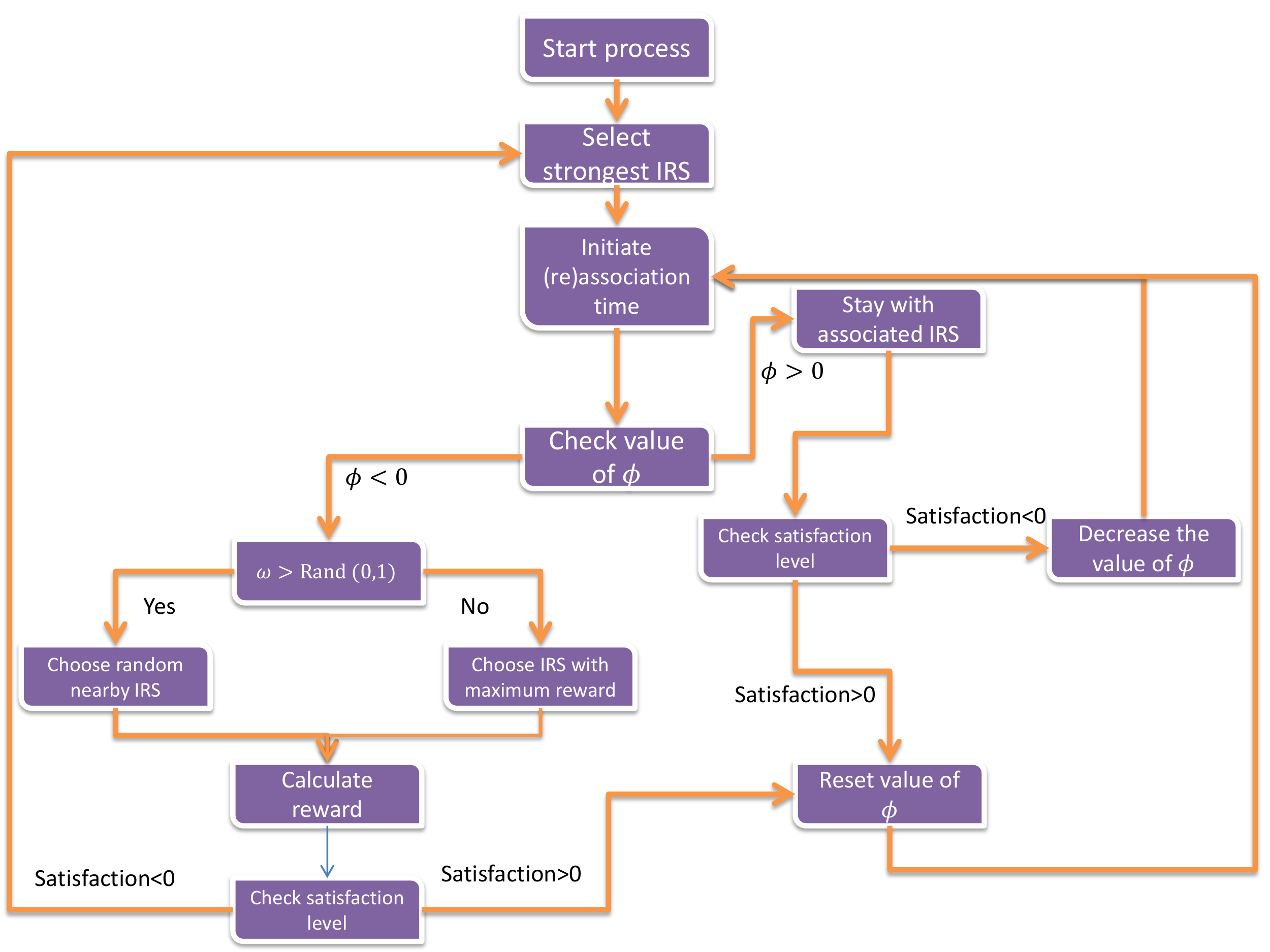}
         \caption{}
         \label{fig:2}
     \end{subfigure}
        \caption{Illustration of (a) ISRE-based 6G vehicular aided HetNet (b)Proposed Contextual Bandit (CB) framework. Here, BS: Base Station, UE: User Equipment, IRS: Intelligent Reflecting Surface.}
        \label{fig:three graphs}
\end{figure*}

In ISRE-based 6G vehicular aided HetNet, one of the most important tasks to prevent information leakage is to associate users with an appropriate IRS array. This would not only reduce the impact of fading but also help in improving the received signal at the intended receiver. Therefore, this section presents a Contextual Bandit (CB) approach where the User Equipment (UE) explores different IRSs in its coverage region and selects the best that satisfies the security requirements. 
\subsection{Network Setup}
Let us consider a downlink vehicular aided HetNet with single macro-cell BS and multiple small-cell BSs as Fig. \ref{blocks} (a) shows. Each small-cell BS is surrounded by multiple IRSs arranged circularly around the edge of the cell. The small-cell is assumed to operate at 5-GHz and it is connected to the macro-cell BS with optical fiber. There are multiple eavesdroppers outside around the IRS along with legitimate vehicular UEs. Each UE is assumed to be equipped with a CB agent for selecting the appropriate nearby IRS for securely receiving the data from small-cell BS. After a certain re-association period, the UEs in the small-cell make a decision either to use the same IRS or select a new one.
We assume that the channel between the small-cell BS and the UE experiences deep fading, and therefore the only viable communication link is through the intermediate IRS. Moreover, since the distance from the small-cell BS to IRSs is the same, therefore the link security is more dependent on the link between IRSs and UEs. During the communication between an IRS and a UE, the eavesdroppers nearby aim to decode the information in a non-cooperative manner. However, by selecting the most secure IRS nearby, the UE prevents the information leakage. It is worth noting that neither the UE nor the IRS is aware of the channel state information of the eavesdropper. This is because it is very difficult to obtain such information in practical networks as eavesdroppers mostly operate passively to hide their existence. In this case, the secrecy rate maximization cannot be performed and the only way is to maximize the rate of the legitimate link. In the following section, we describe how UE can use the CB framework to select the best nearby IRS.
\subsection{CB Framework}
CB is a learning framework that selects the best reward and takes several actions from an action space and makes a series of decisions. As these decisions are taken, the agent must make a trade-off between exploration and exploitation. We represent this tradeoff term as $\omega$ which determines whether the learning agent explores the network or exploits it. Before taking an action, the agent is provided a context or some environment-related information by the BS. After taking an action, the reward is given for the action performed and the goal of the learning agent is to maximize the long-term reward. 
We apply the CB approach to find and select the best IRS for each user to maximize the long-term reward of the UEs in HetNets. The reward of the network is based on the achievable rate of the main link because the rate of the eavesdropper link is unknown in the worst-case scenario. During each association period, if the selected IRS can achieve the desired rate, then the UE is considered to be satisfied. As a result, the reward of the current IRS is increased by one. Otherwise, it does not change and remains zero. During the selection processes, each UE initially selects the IRS with the strongest signal strength. After the initial IRS is selected, the UE then calculates the accumulated reward of all nearby IRSs. If the UE uses the exploitation mode of the reinforcement learning, it selects the IRS with the largest accumulated reward. In contrast, if the UE further explores the environment, it re-associates with a random IRS from the nearby IRS detected. However, if the UE selects the IRS with the largest reward, it remains associated with it for $\phi$ consecutive unsatisfactory periods. Fig. \ref{blocks} (b) describes the CB approach.
\subsection{Results and Discussion}
\begin{figure*}[!t]
\centering
\begin{tabular}{cc}
\includegraphics[scale=.3]{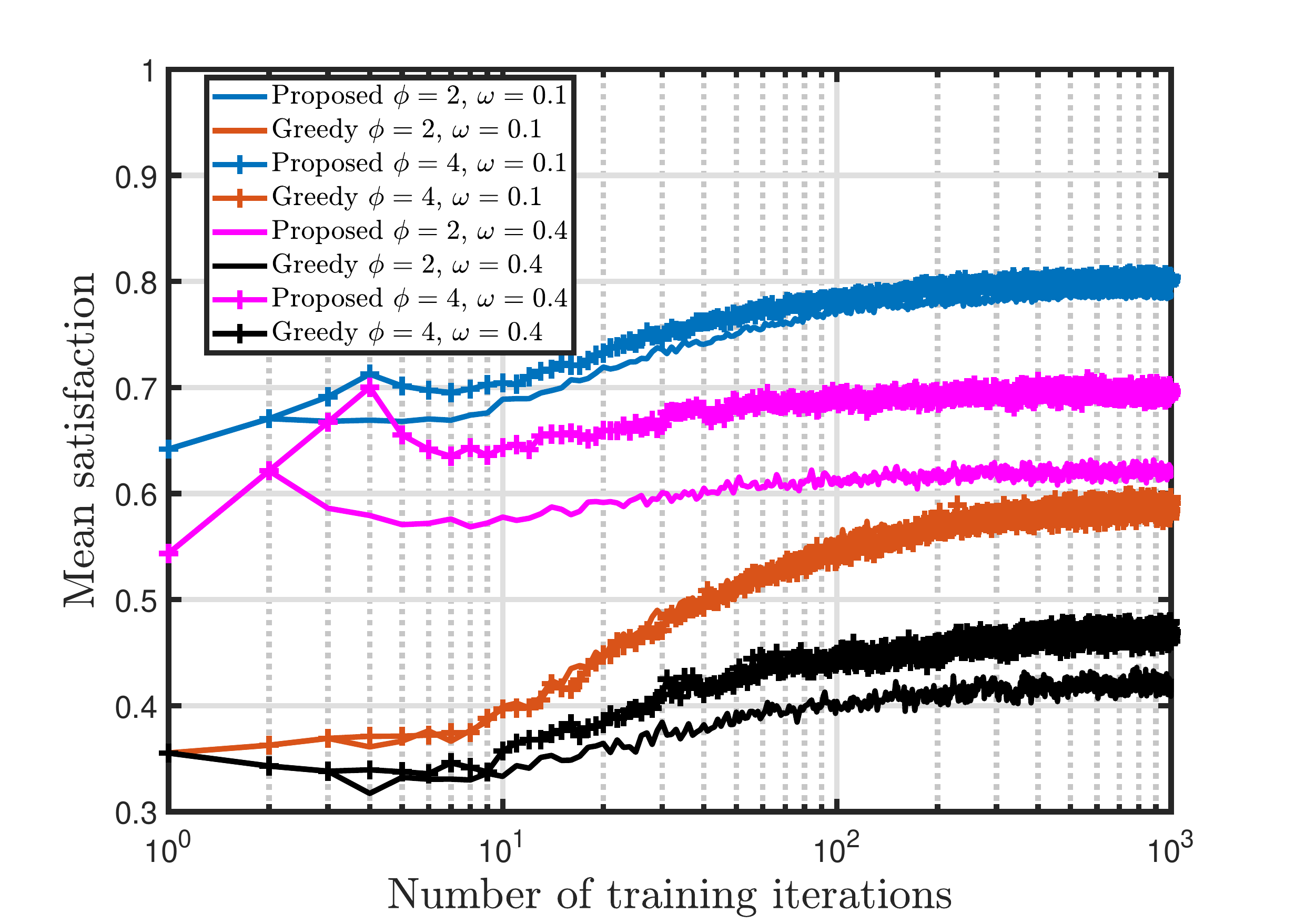} &
\includegraphics[scale=.3]{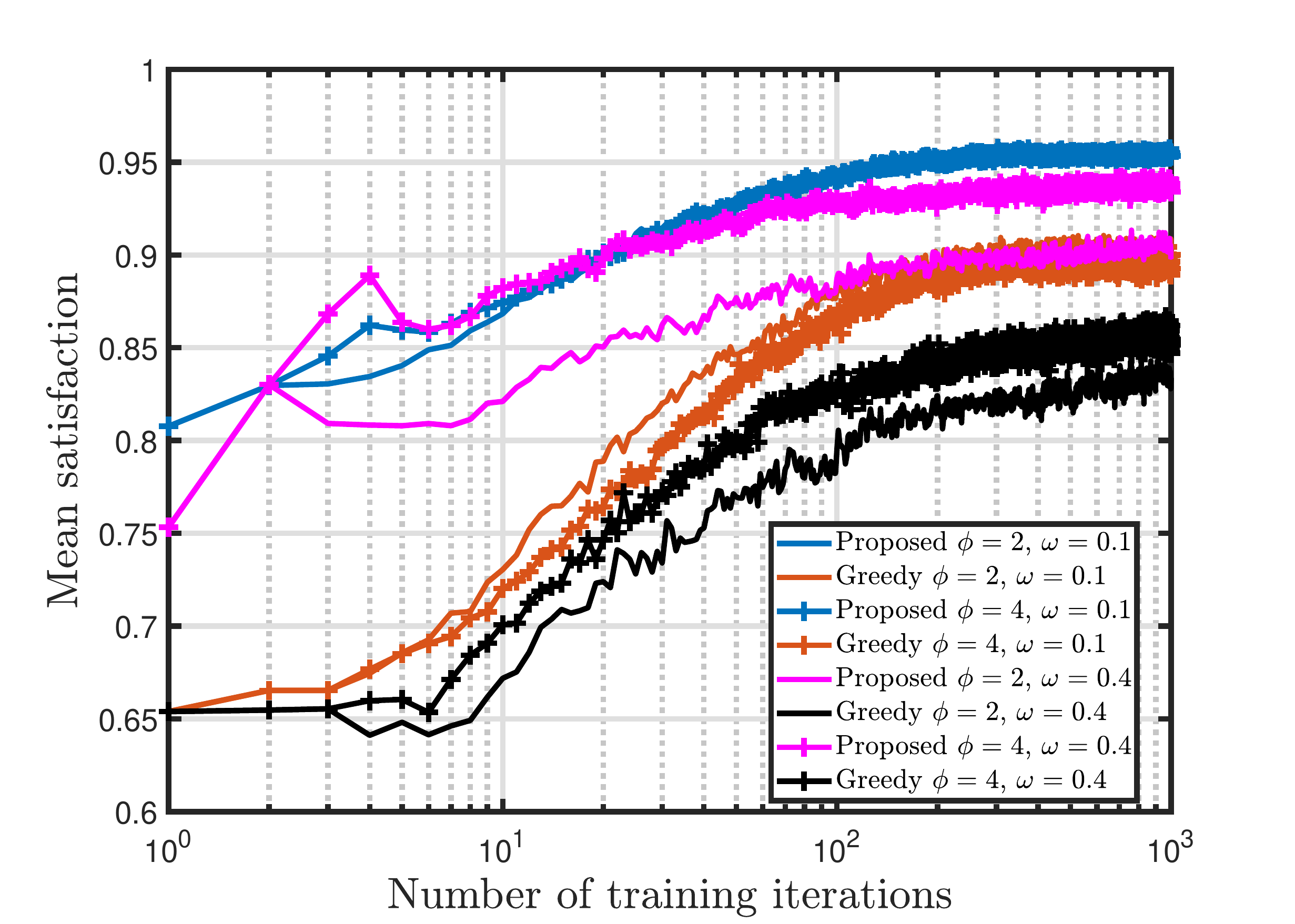}\\
(a) & (b)
\end{tabular}
\caption{Mean satisfaction for (a) Case 1: Random distribution of UEs (b) Case 2: Clustered distribution of UEs.}
\label{figresult}
\end{figure*}

Next, we present and discuss the results obtained based on the extensive Monte-Carlo simulations in MATLAB. The main performance metric is the mean satisfaction which refers to the satisfaction of UEs over the total number of UEs in the network. We place a single macro-cell BS at the center of a square grid along with two small-cell BSs in the network. The positions of all the IRSs are considered fixed and are placed at a radius of 20 meters around the small-cell BS. To generate the simulation results, we have used a fixed transmit power (i.e., 5 dB), and we considered a total of $10^4$ channels. We compare the proposed approach with greedy technique for association. Specifically, in the case of a greedy approach, the learning agent associates with a random IRS. The agent then selects the IRS with the largest reward if it exploits the environment. For the distribution of UEs, we considered two cases, i.e., Case 1: random distribution and Case 2: clustered distribution. The random distribution case distributes the UEs according to a uniform distribution whereas in the clustering case, the UEs are grouped into a cluster of 10 and the location of the cluster is randomly distributed. 
Fig. \ref{figresult} shows the mean satisfaction as a function of the number of training iterations. Fig. \ref{figresult} shows a comparison between the proposed CB approach and the conventional greedy technique (a) for randomly distributed UEs. We observe that the proposed approach outperforms the conventional approach in terms of mean satisfaction. However, the best performance is achieved when the UE remains associated with an IRS for longer unsatisfactory periods. Moreover, the overall mean satisfaction increases when the total number of iterations increases for Case 1. (b) Fig. \ref{figresult} shows a similar trend for the mean satisfactions for a clustered distribution of UEs. In this case, we note that the overall mean satisfaction reduces but the proposed CB approach still performs better than the greedy approach. In addition, the differences between the curves remain almost unchanged which shows that when the number of iterations increases, there is little to no impact on the mean satisfaction (Case 2).

\section{Challenges and Open Issues}
In this section, we discuss some open research challenges on the integration and feasible adoption of ISRE-based 6G vehicular aided HetNets. The aim of this discussion is to provide future research directions to existing and new researchers working in this field. 
\subsection{Data Streamlining}
Although deep learning models outperform conventional optimization approaches most of the time, they still require a massive amount of high-quality data. Training a model for securing a large and complex 6G vehicular aided HetNet communication architecture requires very high-quality data because they have many parameters for learning. In practical conditions, ISRE-based 6G vehicular aided HetNets require a considerable amount of data and mature streamlining platforms for efficient implementations. 

\subsection{Scalability of ISRE}
One of the biggest challenges for ISRE-based 6G vehicular aided HetNets is scalability. It is well-known that end-to-end deep learning experiences an exponential complexity when the size of the network increases. Recent studies on the autoencoder approach suffer from the ``curse of dimensionality’’. One of the exciting approaches can be the deep unfolding (combination of optimization with deep learning) of existing communication schemes and algorithms \cite{9812512}. In fact, this approach can improve existing signal processing algorithms and techniques by leveraging side information (such as state of the transmission channel). Reduced training and model complexity improve the scalability of these models and will help future research efforts on securing 6G vehicular aided HetNets.

\subsection{Spatial Pattern of IRS}
In a HetNet communication scenario, we expect that the IRS would be deployed on the surface of environmental objects. In a practical setting, such a random configuration of the IRS may result in undesired spatial patterns. Therefore, ISRE-based 6G vehicular aided HetNets are going to be jointly influenced by several IRSs operating under the same network. The combined impact of all these passive IRSs working to improve the link security of 6G vehicular aided HetNets is yet unknown and needs further investigation.

\subsection{Mobility and Interference Management}
Mobility is one of the least developed aspects of IRS. However, it has been extensively explored in PLS studies. From the perspective of 6G vehicular aided HetNets, few works have investigated mobility management, and, to the best of our knowledge, no work has yet been done on mobile eavesdroppers operating in smart environments. Another important opportunity for future research is interference management. For an ISRE-based 6G vehicular aided HetNet, interference management is hard to achieve because the passive IRS may not be able to directly communicate with the other surrounding devices which could cause high interference in the network while ensuring link security.

\subsection{Accurate Physical Models}
Due to the signal cancellation capabilities and signal magnification of the IRS, an ISRE-based channel is expected to exchange a large amount of data throughout the network. Thus, the inclusion of the IRS in the network and the ability to control different elements of the radio environment need to be re-examined from a secrecy capacity perspective. Furthermore, to better understand the performance limits, fundamental theories on channel capacity and scaling laws for ISRE-based 6G vehicular aided HetNets need to be derived and validated empirically. 

\subsection{Hardware Impairments}
Hardware impairments significantly affect the performance of any wireless network. This weakness of devices can become a major issue for programmable passive IRS. The involvement of inexpensive and inefficient meta-surfaces along with incompatible third-party components can severely hinder the realization of ISRE-based 6G vehicular aided HetNets.

\section{Conclusion}
Secure and intelligent 6G vehicular aided HetNets have several applications and use cases for wireless networks. In this work, we have presented a novel ISRE-based approach for 6G vehicular aided HetNets. To this end, we have described enabling technologies along with different features and design goals. We have also presented different applications of ISRE-based 6G vehicular aided HetNets and detailed the opportunities provided by such networks. Subsequently, the results obtained in the case study show that the CB approach outperforms the conventional greedy method and shows the feasibility of ISRE-based 6G vehicular aided HetNets. The challenges outlined and the open issues have revealed that there is a huge potential for future research opportunities. We hope that the results provided here will serve as a strong foundation for future studies in this area. 

\ifCLASSOPTIONcaptionsoff
  \newpage
\fi
\bibliographystyle{IEEEtran}
\bibliography{References}
\begin{IEEEbiographynophoto}{Wali Ullah Khan} (Member, IEEE) (waliullah.khan@uni.lu)
received his PhD degree in Information and Communication Engineering (funded by prestigious Chinese Government Scholarship) from Shandong University Qingdao, China. Currently, he is working with University of Luxembourg, Luxembourg as a research associate.  
\end{IEEEbiographynophoto}

\begin{IEEEbiographynophoto}{Muhammad Awais Javed} (Senior Member, IEEE) (awais.javed@comsats.edu.pk) received the Ph.D. in Electrical Engineering from The University of Newcastle, Australia in Feb. 2015. He is currently working as an Associate Professor at COMSATS University Islamabad, Pakistan. 
\end{IEEEbiographynophoto}

\begin{IEEEbiographynophoto}{Sherali Zeadally} (szeadally@uky.edu) is a Professor at the University of Kentucky. His research interests include Cybersecurity, privacy, Internet of Things, computer networks, and energy-efficient networking. He is a Fellow of the British Computer Society and the Institution of Engineering Technology, England.
\end{IEEEbiographynophoto}

\begin{IEEEbiographynophoto}{Eva Lagunas} (Senior Member, IEEE) (eva.lagunas@uni.lu)
received the Ph.D. degrees in telecommunications engineering from the Polytechnic University of
Catalonia (UPC), Barcelona, Spain, in 2014. she joined the Interdisciplinary Centre for Security, Reliability and Trust (SnT), University of Luxembourg, Luxembourg, where she currently holds
a Research Scientist position.
\end{IEEEbiographynophoto}

\begin{IEEEbiographynophoto}{Symeon Chatzinotas}(Senior Member, IEEE) (symeon.chatzinotas@uni.lu) received the Ph.D. degrees in electronic engineering from the University of Surrey, Guildford, U.K., in
2009. He is currently a Full Professor or the Chief Scientist I and the Co-Head of
the Interdisciplinary Centre for Security, Reliability
and Trust, SIGCOM Research Group, University
of Luxembourg. 
\end{IEEEbiographynophoto}
\end{document}